\documentclass[12pt]{article}
\usepackage{latexsym}
\usepackage{epsfig}
\usepackage{graphicx}
\usepackage{epstopdf}
\usepackage[english]{babel}
\usepackage{epsfig}
\newcommand \jp {$J/\psi \,$}

\input colordvi

\begin{document} 
\title{Neutron tagging of quasielastic \jp photoproduction off nucleus in 
ultraperipheral heavy ion collisions at RHIC energies} 

\author {
M. ~Strikman\\
\it Department of Physics,
Pennsylvania State University,\\
\it University Park, PA  16802, USA\\
M. Tverskoy and M.~Zhalov\\
\it St. Petersburg Nuclear Physics Institute, Gatchina, 188300
Russia}
\date{}
\maketitle
\begin{abstract}
We compare the coherent and quasielastic \jp photoproduction in the peripheral
heavy ion collisions in kinematics of RHIC. Our improved estimate of the total coherent
cross section is a factor of two smaller than the earlier ones. 
We find that the counting rate of quasielastic \jp photoproduction tagged by
neutrons emitted due to cascading of the recoiled nucleon within the residual
nucleus exceeds the rate of  the coherent events. 
We argue that measurements of this process can be used 
to learn about the dynamics of color dipole-nucleon interactions in nuclei
in the wide effective
range of energies of $\gamma A$ interactions.
\end{abstract}

The impressive program of the studies of hard
high energy  electroweak  interactions with nucleons
is being performed at HERA for about ten years. Unfortunately,
it will not be followed by the corresponding program of 
studies of the interactions with
nuclei in the near future.

In this situation it is important to 
continue
at least some of these studies (see for review Refs.\cite{baur}, \cite{acta}) 
using possibilities which become available 
at the heavy ion colliders due to  ultraperipheral  collisions (UPC) of nuclei 
in which nuclei pass at impact parameters $b> 2R_A$.

Recent observation of coherent 
$\rho$-meson production in the UPC of nuclei at 
RHIC \cite{rostar}
has demonstrated the
feasibility of this approach. The measured cross section of the coherent $\rho$
meson photoproduction agrees well with theoretical predictions 
(see, for example, Refs. \cite{rhopred1},\cite{rhopred2},\cite{Goncalves:2005yr}). 
It was also
demonstrated experimentally that a noticeable fraction of the coherent events
is followed by the Coulomb induced neutron emission.
So far,
both the experimental and the theoretical studies were focusing on  coherent
production of vector mesons (VM).
Though the coherent production of the vector mesons  is rather easy to identify
by selecting events where transverse momentum of the VM is sufficiently 
small, $\le  \sqrt{3}/R_A$, it is very difficult to determine  whether a left or right moving nucleus 
was  the source  of the photon which converted into a VM.
 Since the photon flux strongly decreases with 
an increase in the photon energy,  this makes it very difficult to study the
photoproduction of the VM at $s_{\gamma N}> m_V\sqrt{s_{NN}}$.
The only currently proposed idea in this direction \cite{baltz} is to use the properties of the final 
states - the pattern of the emission of the neutrons due to one-side or mutual dissociation of the
 nuclei by their Coulomb fields.  Since the breakup happens 
at smaller impact parameters, one could in principle combine different data sets to separate 
 the higher energy and lower energy contributions.  The effect is rather  weak for
 the  RHIC 
 energies to be of practical use, but may be promising for the LHC.

At the same time, there exists another process of vector meson  production with comparable
cross section
which is sensitive to the dynamics of the VM interaction within the nuclear medium 
- quasielastic  (QE) production, 
$\gamma +A \to V +A'$.  The A-dependence of this process varies from $\propto 
A$ for the case, when absorption is small, to $A^{1/3}$ for the case of strong absorption.   
Thus, the sensitivity to the change of  $\sigma_{VN}$ is as large 
$\propto A^{2/3}$ as for the coherent process\footnote{Note that even small 
objects can be absorbed strongly at sufficiently high energies
due to the leading twist shadowing and due to the higher twist effects.}.

An important feature of most of the current detectors is that it is much easier 
to trigger on the VM production if it is accompanied by a breakup of at least one 
of the nuclei, leading to production of one or more neutrons with energy $\sim E_N$ 
 which hit
a zero degree calorimeter (ZDC). 
The current measurements and numerical estimates indicate that 
at the RHIC energies such excitations in the 
case of, say, coherent  \jp production occur with a probability $\sim (30\div 50)\%$
\cite{baltz}.
At the same time, the removal of a nucleon from a heavy
 nucleus in the quasielastic process
should lead to a significant breakup of the nucleus, resulting in the production of neutrons
with a probability of the order one.
 Hence, the rates of detection of the quasielastic and coherent processes
 in heavy ion scattering  at RHIC
 should be comparable.
 
 An attractive and useful feature of quasielastic photoproduction in 
peripheral heavy ion UPC at colliders
  is that the
   neutrons are  emitted from the nucleus interacting with 
the photon. Then  it will  be straightforward to
resolve an ambiguity between left and right  moving emitters in this case.
As a result, we suggest that, provided the detectors have a good acceptance 
for rapidities away from $y=0$,
 it would be possible to study the dynamics of VM production off nuclei in 
 the QE scattering
in  the significantly wider energy range at RHIC and especially at LHC 
than for the coherent case. 

In this paper we start an investigation of the characteristics of the QE 
processes relevant for their identification in UPC. 
As a starting point, we will use the process of the \jp photoproduction at RHIC energies.
There are several reasons for this choice. Firstly,  at the RHIC energies effective
 \jp absorption in the nuclear medium is small,
  so it is reasonable to use for the initial modeling 
the impulse approximation for the interaction of the photon with individual nucleons for
the process of the incoherent \jp photoproduction off nucleus. 
Secondly,  this process is currently being 
analyzed by the PHENIX collaboration \cite{White}.
On top of this,
mechanisms of the \jp photoproduction and interaction of the \jp with nucleons have been the
  subject of theoretical and
experimental studies  for a long time.  
Experimental studies of \jp photo and 
electroproduction at HERA \cite{hera} found that many features of this process are 
correctly predicted by perturbative QCD, in particular, the energy dependence of the cross section, 
weak $Q^2$ and 
energy  dependence of 
the slope parameter 
of the electroproduction amplitude. 
Also, the QCD inspired models describe reasonably the absolute cross section of the
process. 
However very little is known (\cite{slac},\cite{Sokoloff}) about dependence of the 
coherent and quasielastic photoproduction  of \jp on atomic number which provide 
sensitive tests of the color transparency effects in 
 the propagation of small color
dipoles through the nuclear media.

The cross section of \jp coherent production in UPC at RHIC and LHC was estimated in
a number of papers (see, for example, Refs.\cite{rhopred1},\cite{Goncalves:2005yr},\cite{FSZjpsi})
 with the conclusion that the rates should be sufficiently  high to study the process.
 
 We will argue below that the incoherent \jp photoproduction followed by nuclear breakup
 initiated by the recoiled nucleon
 has a tagging efficiency due to the registering the neutrons by ZDC close 
to one. As a result, we find that 
 the counting rate for this process at RHIC should  be about the same or even exceed the rates
 for coherent processes with the nuclear breakup by the Coulomb field.
 We will demonstrate that, on average, about four neutrons should be emitted 
after the knocked out nucleon escapes though the nuclear media. 
Consequently, analysing the ZDC signals it will be possible to single out QE 
events and identify which of the nuclei was the source of the photons that provides a 
possibility to measure the energy dependence of \jp production up to  
significantly higher
energies.

The purpose of this paper is to provide a quantitative proof of this idea. 
In the following publications we will address a number of issues related to
combining effects of nuclear breakup due to the QE and the Coulomb  mechanisms, attenuation
effects, etc.

The cross section of the photoproduction
of \jp in the peripheral ion-ion collisions in the well known Weizsacker-Williams
approximation\cite{ww}
is given by expression

\begin{equation}
{d \sigma(AA\to V AX)\over dydt}={N_{\gamma}(y)}  {d\sigma_{\gamma A\rightarrow V X}(y,t)\over dt}+
{N_{\gamma}(-y)}  {d\sigma_{\gamma A\rightarrow V X}(-y,t)\over dt},
\label{base}
\end{equation}
where $y=ln {\frac {2\omega} {M_V}}$ is the rapidity and
${N_{\gamma}(y)}$ is the flux of the equivalent photons produced by the 
Coulomb field of the relativistic heavy ion. This quantity can be calculated with a reasonable
precision using  
a simple expression\cite{baur}:
\begin{equation}
N(y))=\frac {Z^2\alpha} {{\pi}^2}\int d^2b \Gamma_{AA}({\vec b}) \frac{1} {b^2}X^2
\bigl [K^2_1(X)+\frac {1} {\gamma} K^2_0(X)\bigr ].
\end{equation}
Here $K_0(X)$ and $K_1(X)$ are modified Bessel functions with
argument $X=\frac {bM_Ve^y} {2\gamma}$, $\gamma$ is the Lorentz factor and 
${\vec b}$ is the impact parameter. 
The Glauber profile factor, 
\begin{equation}
\Gamma_{AA}({\vec b})=exp\biggl (-\sigma_{NN}
\int \limits^{\infty}_{-\infty}dz\int d^2b_1
\rho_A(z,{\vec b_1})\rho_A(z,{\vec b}-{\vec b_1})\biggr ),
\end{equation} 
accounts for the inelastic strong interactions of
the nuclei at impact parameters  $b \le 2R_A$ and, hence, suppresses
the corresponding contribution to the vector meson photoproduction.

As a first step we consider the incoherent \jp photoproduction 
off the nucleus, neglecting the initial and final state interaction
of the $c\bar c$ wave package which evolves into a 
 quarkonium with the residual nucleus. 
According to our estimate in Ref.\cite{oscil} the total \jp N cross section 
is on the level of $3.5\pm 0.5$ mb. A similar number comes from a consideration of the
$q\bar q - nucleon$ cross section for the energies of RHIC and transverse sizes 
characteristic for \jp photoproduction.
Hence,  use of the impulse approximation 
seems to be quite reasonable. At the same time the dependence of the 
elementary amplitude $\gamma +N\to J/\psi +N$ on the momentum transfer $t$ is rather flat -
the slope parameter $B_{J/\psi N}
\sim 4$ GeV$^{-2}$. Hence, the effective range of $t$ in quasielastic production
can be rather large,
up to 1 GeV$^2$, as compared to the case of  coherent \jp photoproduction  off
nuclei where the relevant values of $t$ are $-t\le 0.015$ due to suppression of 
higher momentum transfer by the nuclear formfactor. 
The probability to break up
the nucleus by the recoiled nucleon
with momentum  $p_{N}\approx \sqrt{(-t)}\le 1$ GeV in the nucleus rest frame 
is high enough because of the large
total nucleon-nucleon
cross section for this range of momenta.

\begin{figure}
\centering
        \epsfxsize=0.75\hsize
       \epsffile{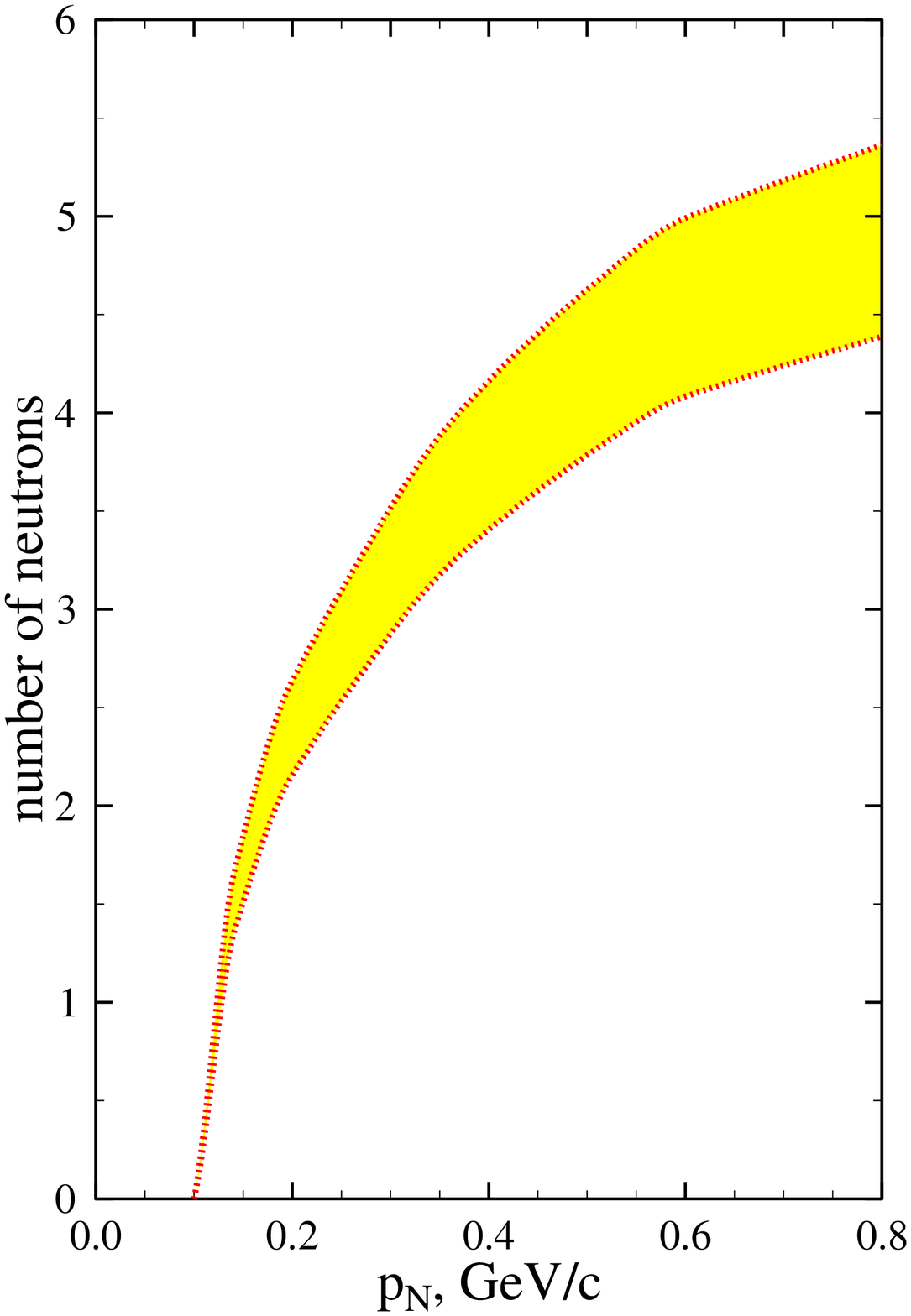}
\caption{
Average number of neutrons for the incoherent production of 
$J/\psi$ 
in UPC of Au at RHIC as a function of the recoiled nucleon momentum $p_{N}=\sqrt{-t}$.}
\label{neutmom}
\end{figure}

To characterize the process of the interaction of the recoiled nucleon with the
residual nucleus in the reaction
$N+(A-1)\to C_{i}+kn$ we introduce  the excitation function $\Phi_{C_{i},kn}(p_N)$
which is the probability to produce exactly $k$ neutrons and any 
number of the charged fragments  $C_{i}$. 
The excitation function $\Phi_{C_{i},kn}(p_N)$ has been calculated using the Monte-Carlo
code accounting for the cascading of the nucleon within the nuclear medium followed by
the evaporation of nucleons and fragments. 
In Ref. \cite{window} we used the same Monte Carlo code to analyze the neutron
 production in the fixed target 
experiment at FNAL (E665) which studied soft neutron production
in deep inelastic scattering of muons off lead. We obtained 
a good description of these data
 \cite{E665} as well as of
 the various data on production of neutrons in the proton -nucleus scattering
at intermediate energies. 
The dependence of the average number of the emitted neutrons from the residual
nucleus on the momentum of the recoiled nucleons is shown in Fig.\ref{neutmom}.
One can see from the figure that for typical \jp transverse momentum in the QE process
$\sim B^{-1/2}_{J/\psi N}\sim .5 \, GeV/c$ in average about four neutrons per event
should be emitted.

The  
 cross section of the incoherent \jp photoproduction accompanied by the breakup
of the residual nucleus is given 
by the expression
\begin{eqnarray}
{d\sigma \over {dt\,dy}}=A\cdot {d\sigma_{\gamma +N\to J/\psi +N}(s,t)\over dt}
\cdot \sum
\limits_{C_{i},k}^{}
\Phi_{C_{i},kn}(p_{N}),
\end{eqnarray}
where $s=2\gamma m_{J/\psi}\exp(y)$ and the momentum transfer 
$-t={m_{J/\psi}^4m_{N}^2/{s}^2}+t_{\bot }$.

\begin{figure}
\centering
        \epsfxsize=0.75\hsize
       \epsffile{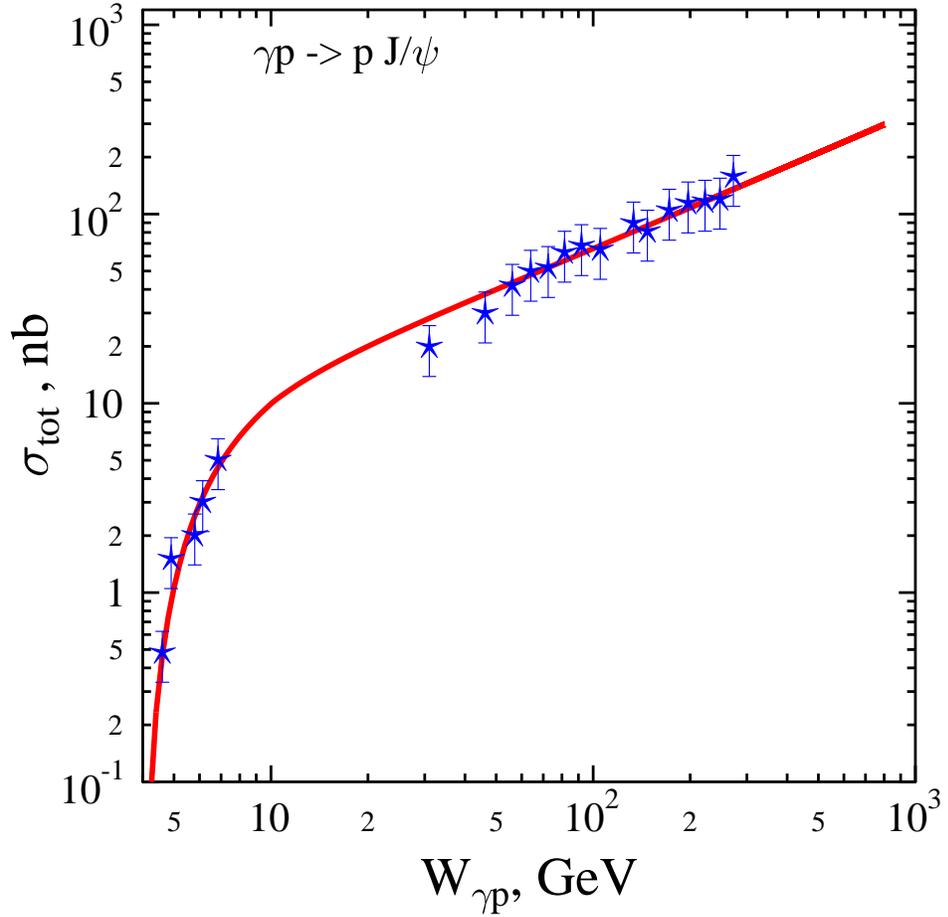}
\caption{
The total cross section of the 
$\gamma +p-> J/\psi +p$ production  
as a function of the $W_{\gamma p}=\sqrt{s_{\gamma p}}$.
 Experimental data from \cite{H1}, solid line - fit to the data using the
parametrization of cross section given by Eq.\ref{eq:cs}. }
\label{eltotcs}
\end{figure}

The cross section of photoproduction off nucleon was parametrized 
by the QCD motivated formula with free parameters fitted  
 to the existing data \cite{H1}:
\begin{eqnarray}
 {d \sigma_{\gamma N\to J/\psi N}(s,t)\over dt}=280\cdot 
\biggl [1-\frac {(m_{J/\psi} +m_{N})^2} {s}\biggr ]^{1.5}\cdot
\biggl ({s\over 10000}\biggr )^{0.415} 
\nonumber \\ 
\biggl [\Theta \bigl ({s_{0}-s}\bigr ) \biggl [ 1-{t\over t_{0}}\biggr ]^{-4} +
\Theta ({s- s_{0}}) exp(B_{J/\psi}t)\biggr ].
\label{eq:cs}
\end{eqnarray}
Here $t_{0}=1 $ GeV$^2$ and the slope parameter for $J/\psi\,N$ scattering was 
parametrized by the expression, $$B_{J/\psi}=3.1+0.25\log_{10}(s/s_{0}),$$
with $s_{0}=100$ GeV$^2$.
This fit (Fig.\ref{eltotcs}) gives a good description of all existing data.
 At the same time,  we found that the low energy
 extrapolation of the Landshoff-Donnachie
parametrization of the \jp N cross section\cite{LD} which we have used in our previous
paper\cite{FSZjpsi} to estimate coherent \jp photoproduction off nuclei in the UPC of heavy ions 
significantly overestimates the value of cross section
at rapidities away from zero.
Moreover, from our study of the
low energy coherent photoproduction of \jp off nuclear targets\cite{oscil} we find
that it is  more reasonable to 
use a larger 
effective \jp N cross section in the region of relatively  
large $x\equiv M^2_{J/\psi}/s \ge 0.01$ relevant for RHIC: 
$\sigma_{eff}(x\ge 0.015)=3$ mb compared to the value $1$ mb used in \cite{FSZjpsi}. 
In total,  these
modifications  resulted in the reduction of 
the cross section of the coherent \jp photoproduction off gold in the kinematics of RHIC by the
factor $\approx 2$.  The  coherent and incoherent 
 \jp photoproduction cross sections in UPC, integrated over rapidity and momentum transfer
for kinematics of RHIC, are given in table \ref{tcrsec}.
In this table we also present the partial cross sections of quasielastic 
\jp production without any emitted neutrons in both direction of the collision indexed
by  $(0n,0n)$ and the cross sections with nuclear breakup with the number $X\ge 1$ neutrons 
in one of two directions
$(0n,Xn)$.

\begin{table}[]
\centering
\begin{tabular} {|c|c|c|c|c|}\hline 
 approximation &    coherent  & incoherent & incoherent 0n,0n & incoherent 0n,Xn \\ \hline
 Impulse &   212 $\mu b$   & 264  $\mu b$     &  38 $\mu b$   & 215 $\mu b$  \\ \hline
 Glauber &   168 $\mu b$     & 210 $\mu b$    &   30 $\mu b$ &   170 $\mu b$  \\ \hline
\end{tabular}
\caption{Total cross sections of coherent and incoherent $J/\psi$ photoproduction calculated
in the Impulse and the Glauber approximations for $Au+Au\to Au+X+J/\psi$
 in UPC at RHIC. }
\label{tcrsec}
\end{table}

\begin{figure}
\centering
        \epsfxsize=0.75\hsize
       \epsffile{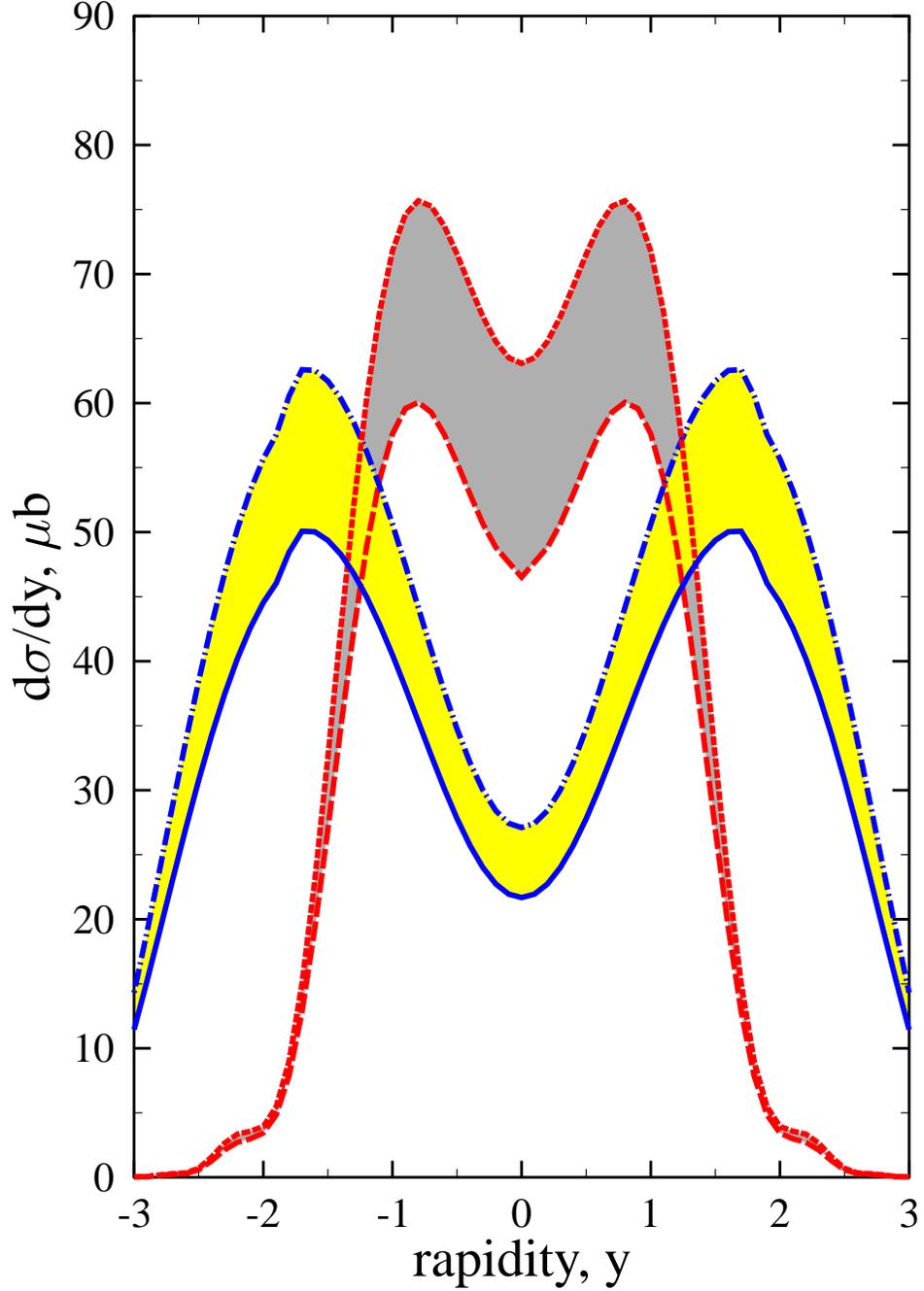}
\caption{
The integrated over momentum transfer rapidity distributions for the 
$J/\psi$  coherent photoproduction 
in UPC of Au ions at RHIC calculated with  effective cross section
for \jp - nucleon interaction of 3 mb
(long-dashed line) and in the Impulse Approximation (short-dashed line)
The incoherent \jp production cross section estimated in the Glauber model (solid line)
and calculated in the IA (dot-dashed line) .}
\label{rap}
\end{figure}

 The rapidity distributions for the coherent and quasielastic
\jp photoproduction, integrated over the momentum transfer, are shown in
Fig.~\ref{rap}. The coherent distribution is more narrow due to  suppression
by the nuclear formfactor in the region where the longitudinal transferred momentum
$p_{z}={m_{J/\psi}^{2}m_{N}\over {s}}$ is still significant.

\begin{figure}
\centering
        \epsfxsize=0.75\hsize
       \epsffile{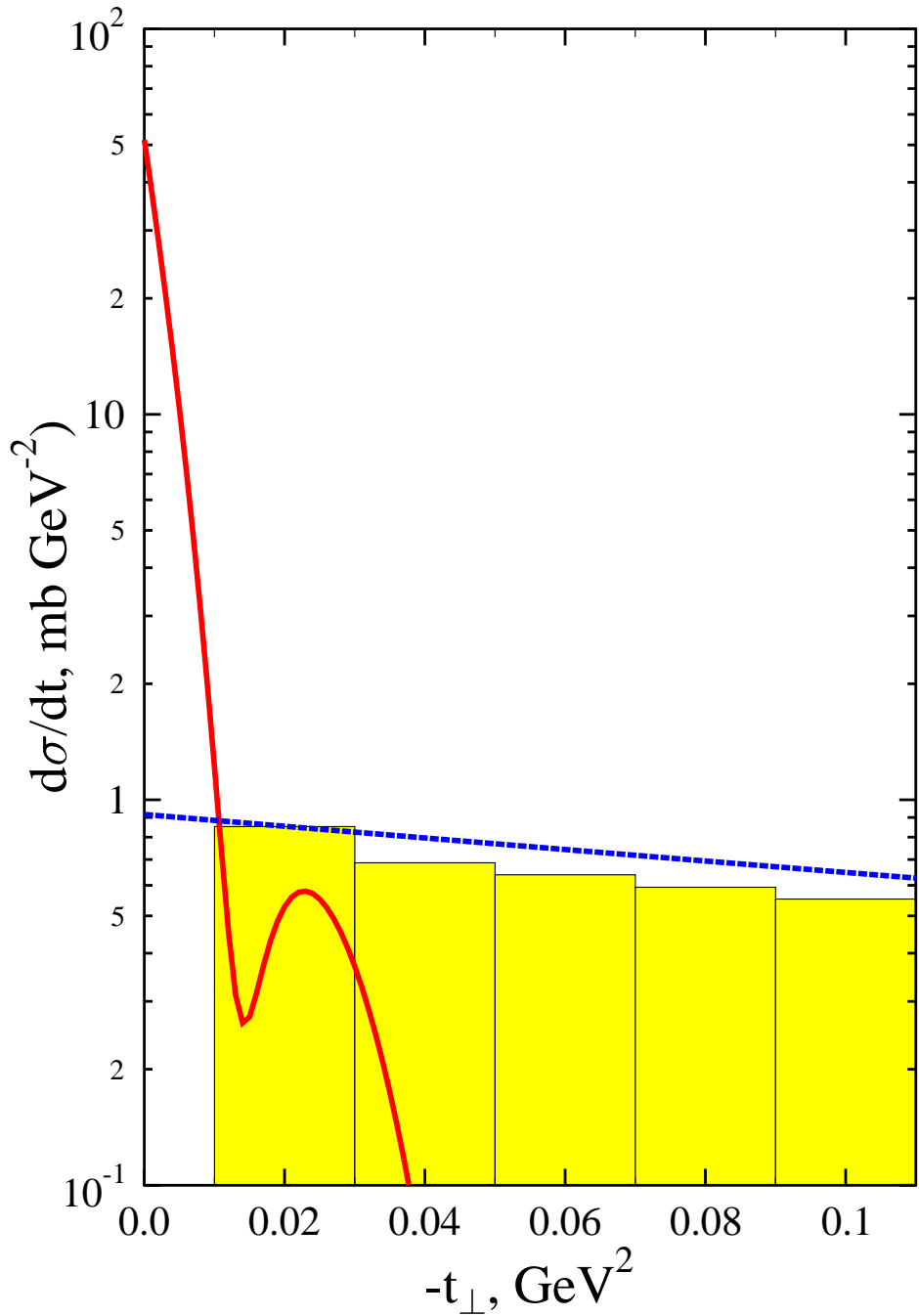}
\caption{Integrated over rapidity ( $-3 \le y\le 3$)
the momentum transfer distributions for the 
$J/\psi$ in the coherent(solid line) and incoherent(dashed line) photoproduction  
in UPC of Au ions at RHIC and the cross section of the incoherent
photoproduction(shadowed hystogram) accompanied by neutrons}
\label{dst}
\end{figure}

\begin{figure}
\centering
        \epsfxsize=0.75\hsize
       \epsffile{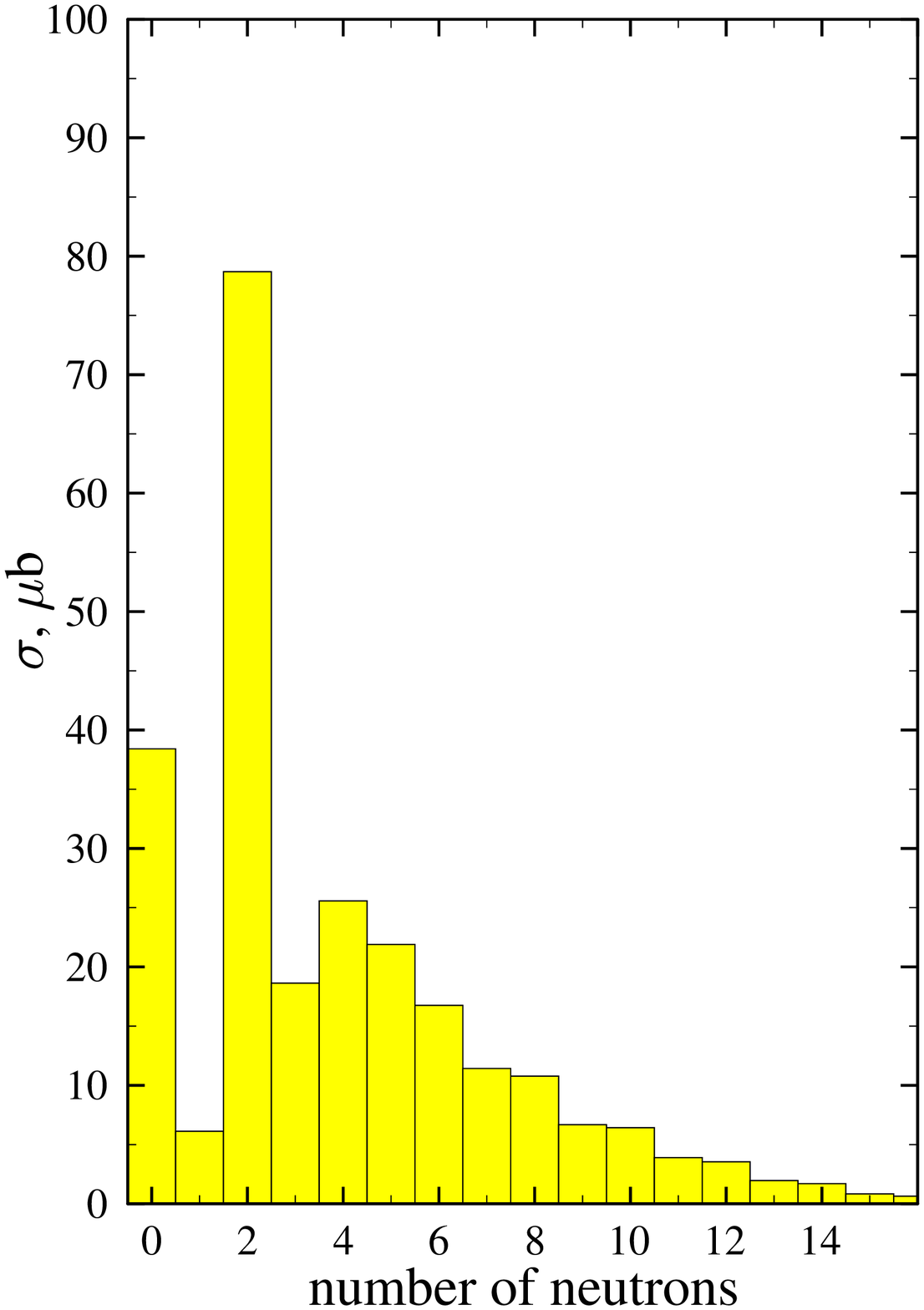}
\caption{
The incoherent cross section for the 
$J/\psi$ production  
in UPC of Au ions at RHIC as a function of the number of accopanied neutrons}
\label{neutdist}
\end{figure}

The dependence of the cross sections, integrated over rapidity,
 on the momentum transfer
are given in Fig.\ref{dst}. It is seen that one can easily discriminate
the coherent and QE events by selecting different transferred momenta.
Actually, at $t\le 0.01$ GeV$^2$ the contribution of the QE  production
(dashed line) is small, however, the QE  mechanism dominates at higher $t$.
The shaded histogram presents the incoherent \jp photoproduction followed by the neutrons, 
emitted due to the final state interaction of the recoiled nucleon. One can see  
that the 
QE  \jp production is accompanied by neutrons with 
a probability very close to one. 
The only exception is 
the  region of very
small momentum transfers  where  the energy of the recoiled particle  
is insufficient to   remove   extra nucleons
 (in gold the minimal separation energy is about 5 MeV). 
Generally, the ratio of the cross section with emission of the
one or more neutrons to the total incoherent
cross section is about 0.8. 
The dependence of the incoherent cross section, integrated over rapidity and momentum
transfer,  on the number of emitted neutrons is presented in (Fig.\ref{neutdist}).
  The distribution has a pronounced peak at multiplicity of neutrons $k=2$ and
the long tail up to the k=14. The average number of the emitted neutrons is $<k_{n}>\approx 4.5$
with $\sqrt {{\sigma}^2}\approx 0.68$.  
  A distinctive feature of the neutron tagging of the incoherent 
  \jp photoproduction is that the \jp produced by the low energy photons
 is moving in the same direction as the emitted neutrons while for the case of \jp production
by the high energy photons the directions of the \jp and neutrons are opposite. 
Namely, this property allows one to select in experimental conditions the incoherent events
of the \jp photoproduction by the high energy photons. 

 Another important point is that, in the first approximation, the Coulomb field induced emission
of neutrons in the coherent process does
not depend on the transverse momentum of \jp. Hence, this
mechanism can be quantified in the coherent production at small $t_{\bot }$ and 
correspondingly  folded in
at higher $t$ in the QE \jp production.
 
The pattern of the neutron emission which we find in QE \jp production
 is qualitatively different from the
case of the electromagnetic excitation. First, according to the prediction of Ref.\cite{baltz} 
a large fraction  of collisions ($\approx 50 \div  70\%$) occur at RHIC
energies without such excitations.  Second, the largest partial channel is the 
emission of one neutron (1n), followed by a two neutron  emission (2n) which constitutes 
about 35\%
of 1n events, and by a long tail
 with a broad and falling distribution \cite{breakup}. On the other hand, in the QE mechanism the 
production of the two neutrons is most likely. 
 Also, there is a different pattern of correlation  between emission
  in two opposite cones in QE and EM mechanism. In the QE mechanism neutrons are 
produced only in one 
of two directions while in the EM mechanism simultaneous
  production  in both directions is possible. 

A more detailed analysis including 
both EM and QE induced neutron emission in quarkonia photoproduction in the UPC 
of ultrarelativistic heavy ions will be presented elsewhere.
  
Note also that in this discussion we neglected diffractive processes of production 
of \jp   with break up of the nucleon: $\gamma + p \to J/\psi
 + M_X$. For relatively small masses, $M_X$, the products will be not detected in the 
 central detector and, hence, the process would be attributed to the QE sample.
 Very little is known about this process at the energies one probes in UPC at RHIC.
 Based on the information at higher energies, one can guess that this process 
should constitute about 10-20\% of the elastic production at $t\sim 0$ and have much 
smaller slope (at least a factor of two smaller). Correspondingly, it will 
further enhance the QE signal.
In principle, it  could be separated  using the t-dependence of the QE \jp production,
as well as the neutron signal.  

In conclusion, we presented an improved estimate of the cross section of 
the coherent \jp photoproduction in UPC at kinematics of the RHIC.
The total cross section is found to be about of 170 $\mu b$ that is approximately
by a factor of  two smaller than the  prediction of  our previous paper \cite{FSZjpsi}
and of  Refs. \cite{rhopred1},\cite{baltz},\cite{Goncalves:2005yr}.
We suggested a new mechanism, the neutron tagging of the incoherent \jp photoproduction
in the UPC of the heavy ions, which can provide a possibility
of reliable selection of the events of \jp production by the high energy photons.

We thank S. White and L. Frankfurt for the discussions. This work was supported 
by the U.S. Department of Energy.

\begin{thebibliography}{99}

\bibitem{baur}
G.~Baur, K.~Hencken, D.~Trautmann, S.~Sadovsky and Y.~Kharlov,
Phys.\ Rept.\  {\bf 364}, 359 (2002),
[arXiv:hep-ph/0112211].

C.~A.~Bertulani, S.~R.~Klein and J.~Nystrand,
  arXiv:nucl-ex/0502005.


\bibitem{acta}
L.~Frankfurt, M.~Strikman and M.~Zhalov,
  Acta Phys.\ Polon.\ B {\bf 34}, 3215 (2003),
  [arXiv:hep-ph/0304301].

\bibitem{rostar}
C.~Adler {\it et al.}  [STAR Collaboration],
Phys.\ Rev.\ Lett.\  {\bf 89}, 272302 (2002),

[arXiv:nucl-ex/0206004].

\bibitem{rhopred1}
S.~Klein, J.~Nystrand,
Phys.\ Rev.\ C {\bf 60}, 014903 (1999),
arXiv:hep-ph/9902259.



\bibitem{rhopred2}
 L.~Frankfurt, M.~Strikman and M.~Zhalov,
  Phys.\ Lett.\ B {\bf 537}, 51 (2002),
  
  [arXiv:hep-ph/0204175];\\
  L.~Frankfurt, M.~Strikman and M.~Zhalov,
  Phys.\ Rev.\ C {\bf 67}, 034901 (2003),
  
  [arXiv:hep-ph/0210303].
 
\bibitem{Goncalves:2005yr}
  V.~P.~Goncalves and M.~V.~T.~Machado,
  arXiv:hep-ph/0501099.

 

\bibitem{baltz}
 A.~J.~Baltz, S.~R.~Klein and J.~Nystrand,
  Phys.\ Rev.\ Lett.\  {\bf 89}, 012301 (2002),
  
  [arXiv:nucl-th/0205031].

\bibitem{White}
S.~White, private communication.

\bibitem{hera}
H.~Abramowicz and A.~Caldwell,
Rev.\ Mod.\ Phys.\  {\bf 71} (1999) 1275,

arXiv:hep-ex/9903037.



\bibitem{slac}
R.~L.~Anderson {\it et al.},
  Phys.\ Rev.\ Lett.\  {\bf 38}, 263 (1977).




\bibitem{Sokoloff}
M.~D.~Sokoloff {\it et al.},
Phys.\ Rev.\ Lett.\  {\bf 57}(1986)3003.




\bibitem{FSZjpsi}
L.~Frankfurt, M.~Strikman and M.~Zhalov,
Phys.\ Lett.\ B {\bf 540}, 220 (2002),

[arXiv:hep-ph/0111221].



\bibitem{ww}
E.~Fermi, Z.\ Physik, {\bf 29}, 315,(1924);\\
C.~F.~von~Weizsacker, Z.\ Physik,\ {\bf 88}, 612,(1934);\\ 
E.~J.~Williams\ Phys.\ Rev.,\ {\bf 45}, 729(1934). 


\bibitem{oscil}
L.~Frankfurt, L.~Gerland, M.~Strikman and M.~Zhalov,
  Phys.\ Rev.\ C {\bf 68}, 044602 (2003).


\bibitem{window}
  M.~Strikman, M.~G.~Tverskoy and M.~B.~Zhalov,
  Phys.\ Lett.\ B {\bf 459}, 37 (1999),
  
  [arXiv:nucl-th/9806099].


\bibitem{E665}
  M.~R.~Adams {\it et al.}  [E665 Collaboration],
  Phys.\ Rev.\ Lett.\  {\bf 74}, 5198 (1995)
  [Erratum-ibid.\  {\bf 80}, 2020 (1998)].




\bibitem{H1}C.~Adloff {\it et al.}  [H1 Collaboration],
Eur.\ Phys.\ J.\ C {\bf 20}(2001)29.


\bibitem{LD}
  A.~Donnachie and P.~V.~Landshoff,
  Phys.\ Lett.\ B {\bf 478}, 146 (2000)
  
  [arXiv:hep-ph/9912312].
\bibitem{breakup}
  M.~Vidovic, M.~Greiner and G.~Soff,
  Phys.\ Rev.\ C {\bf 48}, 2011 (1993);\\
  I.~A.~Pshenichnov, J.~P.~Bondorf, I.~N.~Mishustin, A.~Ventura and S.~Masetti,
  Phys.\ Rev.\ C {\bf 64}, 024903 (2001), [arXiv:nucl-th/0101035];\\
 M.~Chiu, A.~Denisov, E.~Garcia, J.~Katzy and S.~White,
  Phys.\ Rev.\ Lett.\  {\bf 89}, 012302 (2002),
[arXiv:nucl-ex/0109018];\\
  M.~B.~Golubeva {\it et al.},
  Phys.\ Rev.\ C {\bf 71}, 024905 (2005).






\end{thebibliography}
\end{document}